# Improvements and Testing Practical Expressions for Photon Strength Functions of E1 Gamma-Transitions

*Vladimir* Plujko[1,*], *Oleksandr* Gorbachenko[1], *Igor* Kadenko[1], and *Kateryna* Solodovnyk[1]

[1] Nuclear Physics Department, Taras Shevchenko National University, Kyiv, Ukraine

**Abstract.** Analytical expression for the E1 photon strength functions (PSF) is modified to account for the low-energy enhancement due to nuclear structure effects (presence of low-energy state (LES)). A closed-form expression of the E1 PSF function includes response of two nuclear states – LES and giant dipole resonance (GDR). Expression for the nuclear response function on electromagnetic field is based on a model of excitation of two coupled damped states. This approach is tested for different data sets for spherical nuclei. Impact on PSF shape of coupling between LES and GDR excitations is considered.

## 1 Introduction

The average probabilities of $\gamma$-transitions at $\gamma$-ray emission and photoabsorption can be described using the photon (radiative, gamma-ray) strength functions (PSF) [1,2]. These functions are involved in calculations of the observed characteristics of most nuclear reactions. They are also used for investigation of nuclear structure (nuclear deformations, energies and widths of the giant dipole resonances, contribution of velocity-dependent force, shape-transitions, etc.) as well as in studies of nuclear reaction mechanisms. The PSF is important constituent of the compound nucleus model calculations of capture cross sections, $\gamma$-ray production spectra, isomeric state populations, and competition between gamma-ray and particle emission. The most important strength functions in such studies are related to the electric dipole (E1), magnetic dipole and electric quadrupole multipolarities. Different semiphenomenological approaches for the PSF of the E1 transitions were discussed in Refs.[1 - 7].

In this contribution a new closed-form description of the E1 PSF is presented with allowance for electromagnetic response of the two coupled nuclear states - LES and GDR. In spherical nuclei, LES, as a rule, corresponds to the pygmy dipole resonance (PDR). It is most pronounced in spherical atomic nuclei with a large neutron excess. Previously in this case the E1 PSF was approximated by two Lorenzian-like curves. They are associated with response of two independent modes: the PDR mode (oscillations of neutron skin opposite core) and the GDR mode (vibrations in the core of neutrons opposite protons).

Different microscopic and macroscopic approaches indicate on important impact of the relationships between PDR and GDR modes.

We propose a model of two coupled damped state excitation for description of E1 PSF. The calculations within this approach are compared with that for microscopic calculations and available experimental data on photoabsorption cross-sections in the spherical atomic nuclei. The input parameters for the calculations are fixed by the use of the experimental data.

## 2 PSF with allowance for LES response

The photoabsorption cross-section $\sigma_{E1}$ of dipole electric (E1) field is defined by photoabsorption photon strength function $\vec{f}_{E1}$ in the following way:

$$\sigma_{E1}(E_\gamma) = 3E_\gamma (\pi\hbar c)^2 \vec{f}_{E1}(E_\gamma), \qquad (1)$$

where $E_\gamma$ energy of the gamma-rays. For photoabsorption in cold nuclei, the PSF is determined by the imaginary part of nuclear response function $\chi$ on E1 field

$$\vec{f}(E_\gamma) = 8.674 \cdot 10^{-8} \cdot \left(-\frac{1}{\pi} \cdot \operatorname{Im} \chi(E_\gamma)\right), \quad \text{MeV}^{-3}. \quad (2)$$

To take into account the relationships between LES and GDR modes we use nuclear response function of the model of two coupled damped oscillators [8-10] in an external field $E \sim \exp(i\omega t)$:

$$\begin{cases} \ddot{x}_p + (k_p + k)x_p - kx_g + \gamma_p \dot{x}_p = e_p E, \\ \ddot{x}_g + (k_g + k)x_g - kx_p + \gamma_g \dot{x}_g = e_g E, \end{cases} \quad (3)$$

---

[*] Corresponding author: plujko@gmail.com

where $x_{p,g}, \gamma_{p,g}, k_{p,g}$ are displacements, damping widths, restoring forces for nucleons forming low-energy and giant resonances, correspondingly, and the factor $e_\alpha$ ($\alpha = p,g$) determines contribution of state $\alpha$ to response function at action of external field.

As it was shown in Ref. [8], the system (3) can be transformed to the equations

$$\begin{cases} \ddot{y}_p + \omega_p^2 y_p + (\Gamma_p + \gamma) \dot{y}_p - \gamma \dot{y}_g = z_p E \\ \ddot{y}_g + \omega_g^2 y_g + (\Gamma_g + \gamma) \dot{y}_g - \gamma \dot{y}_p = z_g E \end{cases}, \quad (4)$$

by the use of transformation matrix

$$U = \begin{pmatrix} \cos\theta, & \sin\theta \\ -\sin\theta, & \cos\theta \end{pmatrix}.$$

Here, the angle $\theta$ is solution of the equation :

$$\cot^2\theta + \frac{k_g - k_p}{k}\cot\theta = 1,$$

and

$$\begin{pmatrix} y_p \\ y_g \end{pmatrix} = U\begin{pmatrix} x_p \\ x_g \end{pmatrix}, \quad \begin{pmatrix} z_p \\ z_g \end{pmatrix} = U\begin{pmatrix} e_p \\ e_g \end{pmatrix},$$

$$\begin{pmatrix} \Gamma_p + \gamma, & \gamma \\ \gamma, & \Gamma_g + \gamma \end{pmatrix} = U\begin{pmatrix} \gamma_p, & 0 \\ 0, & \gamma_g \end{pmatrix}U^{-1},$$

$$\begin{pmatrix} \omega_p^2, & 0 \\ 0, & \omega_g^2 \end{pmatrix} = U\begin{pmatrix} k_p + k, & k \\ k, & k_g + k \end{pmatrix}U^{-1}.$$

A solution of the system (4) leads to the following analytical expression for the nuclear response function $\chi(E_\gamma)$ on external field with the frequency $\omega = E_\gamma / \hbar$

$$\chi(E_\gamma) = P(E_\gamma; GDR, LES) + P(E_\gamma; LES, GDR) \quad (5)$$

with
$P(E_\gamma; k,l) =$

$$= \frac{z_k^2 + \dfrac{z_k z_l i E_\gamma \gamma}{E_l^2 - E_\gamma^2 + i E_\gamma (\Gamma_l + \gamma)}}{E_k^2 - E_\gamma^2 + i E_\gamma (\Gamma_k + \gamma) + \dfrac{\gamma^2 E_\gamma^2}{E_l^2 - E_\gamma^2 + i E_\gamma (\Gamma_l + \gamma)}}, \quad (6)$$

where $E_m$, $\Gamma_m$ and $z_m$ ($m = k,l \rightarrow LES, GDR$) are the energy, width and contribution of the LES or GDR state; $\gamma$ is parameter of a coupling between two excitation modes. In the case of independent modes ($\gamma = 0$), this expression corresponds to approach with two independent Lorentzians.

The approach with using expression (2) for the PSF with (5) for response function is named below as TSE model in cold spherical nuclei.

In TSE model for hot spherical nuclei, general expressions between response function and PSF of the MLO method [1,2,12] are used; that is, the PSF for photoabsorption ($\vec{f}$) and gamma decay ($\bar{f}$) are calculated by formulae

$$\vec{f}(E_\gamma) = F(E_\gamma, T_i), \quad \bar{f}(E_\gamma) = F(E_\gamma, T_f), \quad (7)$$

where

$$F(E_\gamma, T) = 8.674 \cdot 10^{-8} \left(-\frac{1}{\pi} \cdot \operatorname{Im}\chi(E_\gamma)\right) L(E_\gamma, T), \quad \text{MeV}^{-3}$$

$$L(E_\gamma, T) = \frac{1}{1 - \exp(-E_\gamma / T)},$$

and $T_i$ ($T_f$) is the temperature initial (final) states. Scaling factor $L(E_\gamma, T)$ determines low-energy enhancement of the PSF in heated nucleus and can be interpreted as the average number of 1p-1h states excited by an electric field.

## 3 Calculations and Discussion

We apply the TSE model with LES=PDR for description of the different experimental data for photoabsorption cross-sections. The relationships (1), (2), (5), (6) are used for calculation of the cross-sections. Figures 1, 2 show comparisons of the experimental data with the calculations for isotopes $^{130,132}$Sn [13], $^{88}$Sr and $^{139}$La [14,15].

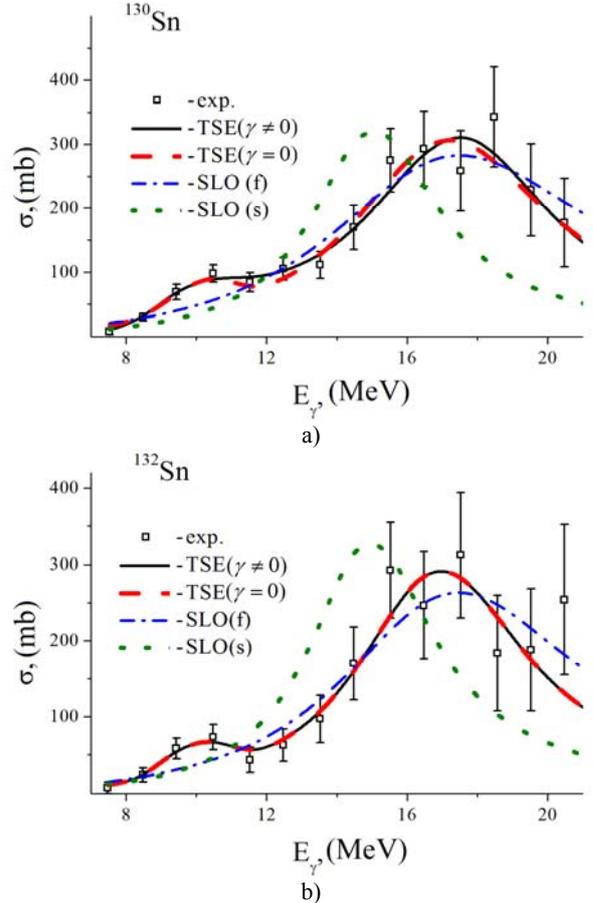

**Fig. 1.** Dependence of photoabsorption cross-section for $^{130}$Sn (a) and $^{132}$Sn (b) on gamma-ray energy. Experimental data are taken from [13] and fitting energy range is $E_\gamma < 21$ MeV.

The TSE expression with zero value of the coupling width ($\gamma = 0$) consists of two independent Lorentzians (for PDR and GDR respectively). The resonance parameters were found by fitting of the experimental data. The initial values of the GDR parameters were taken either from table[4] or from systematics [4,16].

The SLO-curves (SLO(s), SLO(f)) correspond to calculations within a model of standard Lorentzian in spherical nuclei (one Lorentzian with the constant width): GDR parameters in SLO(s) were taken either from Refs. [4,16]; in SLO(f) the parameters were used from fitting experimental data.

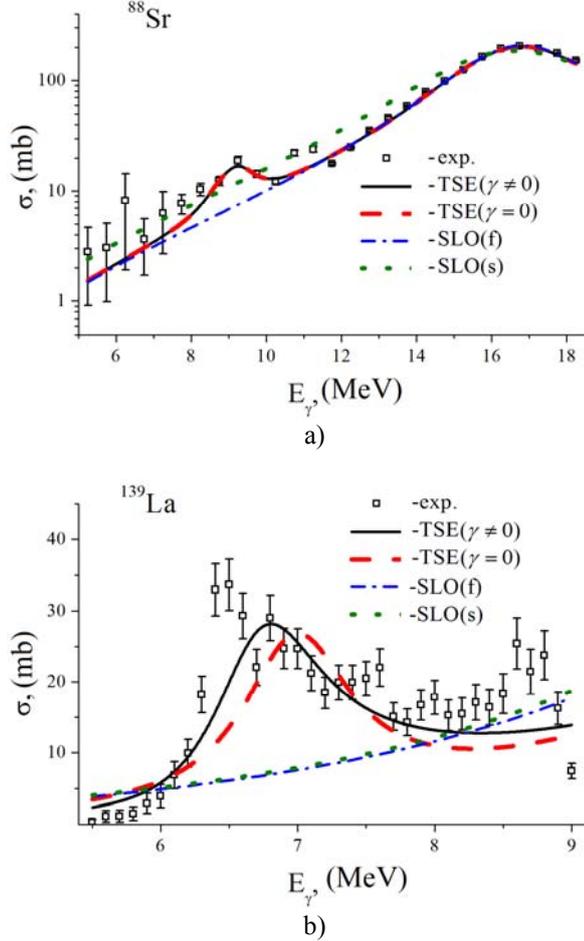

**Fig. 2.** Dependence of the photoabsorption cross-section for $^{88}$Sr(a) and $^{139}$La(b) on gamma-ray energy. Experimental data are taken from [14,15].

The parameters and the least-square deviations ($\chi^2$) in fitting the calculations to data are presented in the Tables 1 and 2, where

$$\chi^2 = \sum_{j=1}^{j_{max}} (\sigma_{exp,j} - \sigma_{th,j})^2 /((j_{max} - n)\cdot \Delta\sigma_{exp,j}^2), \quad (8)$$

$j_{max}$ =14 - number of the experimental data poins of cross-section ($\sigma_{exp,j} = \sigma(E_{\gamma,j})$) at gamma-ray energy $E_{\gamma,j}$ < 21 MeV; $n$ -number of the parameters ($n$ =7 for TSE with $\gamma \neq 0$, $n$ = 6 for TSE with $\gamma = 0$ and $n$ = 3 for SLO). In the tables, parameter $s_\alpha$ is strength of the corresponding resonance (first or second component in Eq.(5)) in units of the Thomas–Reiche–Kuhn sum rule $\sigma_{TRK}$ = 60$NZ/A$ [mb·MeV]:

$$s_\alpha = \int_0^\infty \sigma_\alpha dE_\gamma / \sigma_{TRK}, \quad \alpha = PDR \ or \ GDR.$$

**Table 1.** Comparison of the resonance parameters for $^{130}$Sn.

|  | TSE | TSE | SLO(f) | SLO(s) | [13] |
|---|---|---|---|---|---|
| $\gamma$ | 2.5(13) | 0 | - | - | - |
| $E_{PDR}$ | 9.5(2) | 10.1(2) | - | - | 10.1(7) |
| $\Gamma_{PDR}$ | 0.0(37) | 1.2(11) | - | - | <3.4 |
| $s_{PDR}$ | 0.118(4) | 0.09(4) | - | - | - |
| $E_{GDR}$ | 17.8(5) | 17.2(4) | 17.5(6) | 15.0 | 15.9(5) |
| $\Gamma_{GDR}$ | 3.6(15) | 6.6(12) | 9.4(59) | 4.5 | 4.8(1.7) |
| $s_{GDR}$ | 1.58(25) | 1.76(12) | 2.26(29) | 1.22 | - |
| $\chi^2$ | 0.6 | 1.0 | 2.6 | 8.3 | - |

**Table 2.** Comparison of the resonance parameters for $^{132}$Sn.

|  | TSE | TSE | SLO(f) | SLO(s) | [13] |
|---|---|---|---|---|---|
| $\gamma$ | 0.0(6) | 0 | - | - | - |
| $E_{PDR}$ | 10.0(1) | 10.0(1) | - | - | 9.8(7) |
| $\Gamma_{PDR}$ | 0.5(28) | 0.5(28) | - | - | <2.5 |
| $s_{PDR}$ | 0.09(43) | 0.09(25) | - | - | - |
| $E_{GDR}$ | 17.0(5) | 17.0(5) | 17.4(8) | 15.0 | 16.1(7) |
| $\Gamma_{GDR}$ | 5.5(13) | 5.5(13) | 8.1(17) | 4.4 | 4.7(2.1) |
| $s_{GDR}$ | 1.41(22) | 1.41(26) | 1.83(33) | 1.22 | - |
| $\chi^2$ | 0.61 | 0.54 | 1.4 | 5.4 | - |

The PDR parameters obtained from fitting TSE model with $\gamma = 0$ to data practically coincide with that ones from Ref. [13]. There are some difference in values of obtained GDR energies and results from Ref. [13]. In our opinion, it results from different energy intervals of fitting. For $^{130}$Sn, the values of the least-square $\chi^2$ and GDR width are reduced with allowance for coupling $\gamma$ between PDR and GDR modes. In this case TSE approach provides better description of the experimental data and more accurate determination of GDR parameters. For $^{132}$Sn, coupling between PDF and GDR mode is not manifested.

Fig.3 demonstrates impact of coupling parameters $\gamma$ on shape of photoabsorption cross-sections in $^{130}$Sn. It can be seen that in the GDR range the peak-value of cross-section is sharply decreased and GDR width is increased with increasing coupling parameter $\gamma$.

Table 3 shows the average values of least-square deviation $<\chi^2> = \sum_{i=1}^{N} \chi_i^2 / N$ in fitting the calculations to data for $N$ nuclei: $^{130,132}$Sn [13], $^{88}$Sr, $^{89}$Y, $^{90}$Zr, $^{92,94,96,98,100}$Mo, $^{124,128,134}$Xe, $^{139}$La [14,15], $^{91,92,94,96}$Zr, $^{95,97,98,100}$Mo, $^{105,106,108}$Pd, $^{116,117,118,119,120,122,124}$Sn, $^{139}$La, $^{141}$Pr [17] (these data were taken from EXFOR database).

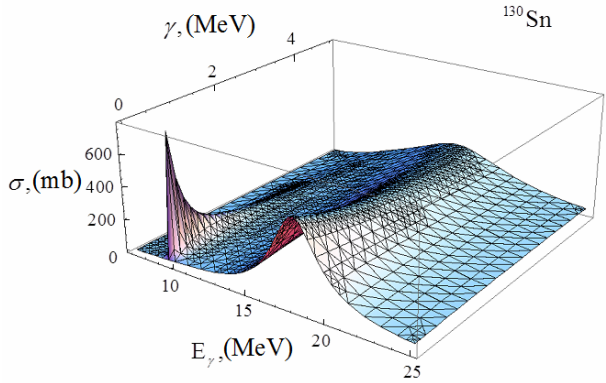

**Fig. 3.** Photoabsorption cross section of $^{130}$Sn versus gamma-ray energy $E_\gamma$ and coupling width $\gamma$.

It can be seen from results presented in the Table 3, the TSE approach with nonzero coupling $\gamma$, provides better description of the experimental data.

**Table 3.** The average values $<\chi^2>$ of least-square deviations of the theoretical calculations from experimental data for $N$ nuclei.

| Main Ref. for data | $N$ | Model | | | |
|---|---|---|---|---|---|
| | | TSE($\gamma \neq 0$) | TSE($\gamma = 0$) | SLO(f) | SLO(s) |
| [13] | 2 | 0.6 | 0.8 | 2.0 | 6.8 |
| [14,15] | 12 | 8.6 | 10.7 | 17.9 | 149.9 |
| [17] | 20 | 8.7 | 13.4 | 14.3 | 224.0 |

Table 4 shows average values $<\chi^2>$ of least-square deviation of the calculations within different phenomenological models from the microscopic results [18] within framework of quasi-particle random phase approximation (QRPA) and the quasi-particle time blocking approximation (QTBA). The microscopic calculations for isotopes $^{58,68,72}$Ni, $^{116,118,122,132}$Sn were used; they were taken from corresponding figures in Ref.[7,18] with energy step 0.01 MeV. The uncertainty $\Delta\sigma_{\exp,j}$ for $\chi_i^2$, Eq.(8), was adopted equal to 10 percent of corresponding microscopic value of $\sigma_j$.

**Table 4.** The average values $<\chi^2>$ of least-square deviations of the phenomenological calculations from microscopic results.

| Ref. for data | $N$ | Model | | | |
|---|---|---|---|---|---|
| | | TSE ($\gamma \neq 0$) | TSE ($\gamma = 0$) | SLO(f) | SLO(s) |
| [18] QRPA | 7 | 7.5 | 15.6 | 26.4 | 2214.6 |
| [18] QTBA | 7 | 8.1 | 20.1 | 34.3 | 653.6 |

It can be seen, that TSE model much better describes the microscopic calculations then SLO approaches.

On the whole, proposed TSE model is a simple approach to account for low energy enhancement due to LES excitation. Allowance for coupling between low and high energy modes leads to better description of the experimental data and microscopic calculations in comparison with situations of independent modes. Therefore, TSE approach can provide more accurate determination of the resonance parameters (both PDR and GDR).

This work is supported in part by the IAEA (Vienna) under IAEA Research Contract within CRP #F41032

# References

1. T. Belgya *et al.*, *Handbook for Calculations of Nuclear Reaction Data: RIPL-2*, IAEA-TECDOC-1506 (IAEA, Vienna, 2006); http://www-nds.iaea.org/RIPL-2/
2. R. Capote *et al.*, Nucl. Data Sheets **110**, 3107 (2009); http://www-nds.iaea.org/RIPL-3/
3. A.R. Junghans *et al.*, Phys. Lett. B **670**, 200 (2008)
4. V.A. Plujko *et al.*, At. Data Nucl. Data Tables **97**, 567 (2011)
5. V. Avrigeanu *et al.*, Phys. Rev. C **90**, 044612 (2014)
6. V.A. Plujko, *et al.*, Nucl. Data Sheets **118**, 237 (2014)
7. O. Achakovskiy *et al.*, Phys. Rev. C**91**, 034620 (2015); https://arxiv.org/abs/1412.0268v1
8. A.S. Barker, J.J. Hopfield, Phys. Rev. **135**, A1732 (1964)
9. V.Baran *et al.*, Rom. Journ. Phys. **57**, 36 (2012)
10. V.Baran *et al.*, Phys. Rev. C **85**, 051601(R) (2012)
11. A.Croitoru *et al.*, Rom. Journ. Phys. **60**, 748 (2015)
12. V.A.Plujko (Plyuiko), Sov. J. Nucl. Phys. **52**, 639 (1990)
13. P. Adrich *et al.*, Phys. Rev. Lett. **95**, 132501 (2005)
14. A. Makinaga *et al.*, Phys. Rev. C **82**, 024314 (2010)
15. R. Schwengner *et al.*, Phys. Rev. C **78**, 064314 (2008)
16. V.A. Plujko *et al.*, Journ. Korean Phys. Soc. **59**(2),1514 (2011)
17. H. Utsunomiya *et al.*, Phys. Rev. C**82**, 064610 (2010)
18. O. Achakovskiy, *et al.*, EPJ Web of Conferences, **107,** 05002(2016); EPJ Web of Conferences, **107,** 05005(2016)